\documentclass[
    ,final            
  ]
  {aipproc}

\usepackage{bm} 
\layoutstyle{6x9}

\newcommand{\sT}{{\scriptscriptstyle T}}

\begin{document}

\title{Accessing the distribution of linearly polarized gluons in 
       unpolarized hadrons}


\author{Dani\"el Boer}{
  address={Theory Group, KVI, University of Groningen,
Zernikelaan 25, NL-9747 AA Groningen, \\The Netherlands  }
}

\author{Stanley J.~Brodsky}{
  address={SLAC National Accelerator Laboratory, Stanford University,
Stanford, California 94309, USA,  CP$^3$-Origins, Southern Denmark University, Odense, Denmark} 
}

\author{Piet J.~Mulders}{
  address={Department of Physics and Astronomy, Vrije Universiteit
  Amsterdam, NL-1081 HV Amsterdam, The Netherlands}
}

\author{Cristian Pisano\footnote{Speaker. Talk given at the XIX Workshop on Deep-Inelastic Scattering and Related Subjects (DIS 2011), April 11-15, Newport News, VA, USA.}  } {
  address={Dipartimento di Fisica, Universit\`a di Cagliari, and INFN, Sezione di Cagliari, I-09042 Monserrato (CA), Italy} 
}

\begin{abstract}
Gluons inside unpolarized hadrons can be linearly polarized provided they have
 a nonzero transverse momentum. 
The simplest and theoretically safest way to probe this distribution of linearly polarized gluons is through $\cos 2\phi$ asymmetries in heavy quark pair or dijet production in electron-hadron collisions.
Future Electron-Ion Collider (EIC) or Large Hadron electron Collider (LHeC) 
experiments
 are ideally suited for this purpose. Here we 
estimate the maximum asymmetries for EIC kinematics.
\end{abstract}

\maketitle


\section{Introduction}

Linearly polarized gluons in an unpolarized hadron, carrying a light-cone momentum 
fraction $x$ and transverse momentum $\bm p_\sT$
 w.r.t.\ to the parent's momentum, are described by the transverse momentum
dependent distribution (TMD) 
 $h_1^{\perp\,g}(x,\bm{p}_{\sT}^2)$ 
 \cite{Mulders:2000sh,Boer:2009nc,Boer:2010zf}. Unlike the quark TMD 
 $h_1^{\perp \, q}$ of transversely polarized quarks inside an
unpolarized hadron (also frequently referred to as
Boer-Mulders function) \cite{Boer:1997nt}, $h_1^{\perp\,g}$ is chiral-even and $T$-even.  
This means it does not require initial or final state interactions (ISI/FSI) to be 
nonzero. Nevertheless, as any TMD, $h_1^{\perp\, g}$ can receive contributions from 
ISI or FSI and therefore can be process dependent, in other words, non-universal, 
and its extraction can be hampered in nonfactorizing cases.
 
Thus far no experimental studies of $h_1^{\perp\,  g}$ have been performed.  
As recently pointed out, it is possible to obtain an extraction of $h_1^{\perp\, g}$ in a 
simple and theoretically safe manner, since unlike
$h_1^{\perp\, q}$ it does not need to appear in pairs \cite{Boer:2010zf}. 
Here we will discuss observables that involve only a single $h_1^{\perp\, g}$ in 
semi-inclusive DIS to two heavy quarks or to two jets, which allow for TMD 
factorization and hence a safe extraction. 
The corresponding hadroproduction processes  run into the problem 
of factorization breaking \cite{Boer:2010zf,Rogers:2010dm}.

\section{Azimuthal asymmetries}

We first consider heavy quark (HQ) production,
$e (\ell)${+}$h(P)$$\to$$e(\ell^\prime)${+}$Q(K_1)${+}$\bar{Q}(K_2)${+}$X$,
where the four-momenta of the particles are given within brackets, and
the heavy quark-antiquark pair in the final state is almost back-to-back in the
plane perpendicular to the direction of the exchanged photon and hadron.
We look at the heavy quarks crea\-ted in the photon-gluon fusion process,
which can be distinguished kinematically from intrinsic charm production; 
e.g., from the $Q \bar Q$ invariant mass distribution.
The calculation proceeds along the lines explained
in Refs.\ \cite{Boer:2009nc,Boer:2007nd}.
We obtain for the cross section
integrated over the angular distribution of the back-scattered
electron $e(\ell^\prime)$:
\begin{eqnarray}
\frac{d\sigma}
{dy_1\,dy_2\,dy\,dx_{\scriptscriptstyle B}\,d^2\bm{q}_{\sT}\, d^2\bm{K}_{\perp}}\,\, =\,\, 
\frac{\alpha^2\alpha_s}{\pi  s
  M_\perp^2}\, \frac{(1+y x_{\scriptscriptstyle B})}{ y^5 x_{\scriptscriptstyle B}}\,  \bigg( A +
 B \,{\bm q}_{\sT}^2\, \cos 2 \phi \bigg)
\,\delta(1 - z_1 - z_2)\, .
\label{BBMPeq:cso}
\end{eqnarray}
This expression involves the standard DIS variables:
$Q^2 = -q^2$, where $q$ is the momentum of the virtual photon,
$x_{\scriptscriptstyle B}= Q^2/2P\cdot q, y = P\cdot q/P\cdot \ell$ and
$s = (\ell + P)^2 = 2\,\ell\cdot P = 2\,P\cdot q/y = 
Q^2/x_{\scriptscriptstyle B} y$.
Furthermore, we have for the HQ transverse momenta $K_{i\perp}^2 = -\bm
K_{i\perp}^2$ and introduced the rapidities $y_i$ for the HQ momenta
(along photon-target direction). We denote the proton mass with $M$ and the heavy (anti)quark
mass with $M_Q$. For the partonic subprocess we have
$p+q=K_1+K_2$, implying $z_1+z_2 = 1$, where $z_i=P\cdot K_i/P\cdot q$.
We introduced the  sum and difference of the HQ transverse momenta, 
$K_\perp = (K_{1\perp} - K_{2\perp})/2$ and
$q_{\sT} = K_{1\perp} + K_{2\perp}$, considering
$\vert q_{\sT}\vert \ll \vert K_\perp\vert$. 
In that situation,
we can use the approximate HQ transverse momenta
$K_{1\perp} \approx K_{\perp}$ and $K_{2\perp} \approx -K_{\perp}$
denoting $M_{i\perp}^2 \approx M_\perp^2 = M_Q^2 + \bm K_\perp^2$.
The azimuthal angles of $\bm{q}_{\sT}$ and  $\bm{K}_\perp$ are denoted
by $\phi_{\sT}$ and $\phi_\perp$ respectively, and $\phi \equiv \phi_{\sT}-\phi_\perp$.
The functions $A$ and $B$ depend on $y, z (\equiv z_2), Q^2/M_{\perp}^2,
M_Q^2/M_{\perp}^2$, and  $\bm{q}_{\sT}^2$.

The angular independent part $A$ is non negative and
involves only the unpolarized TMD gluon distribution $f_1^g$, 
$A \equiv  e^2_Q \, f_1^g (x,\bm{q}_{\sT}^2)\, {\cal A}^{e  g\to e Q\bar Q} \ge 0 $.  
We focus on the magnitude $B$ of the $\cos 2 \phi$ asymmetry, which is 
determined by $h_1^{\perp\, g}$. Namely,
\begin{equation}
B
= \frac{1}{M^2}\,
 e^2_Q\,  h_1^{\perp \,g} (x, \bm{q}_{\sT}^2)\,
{\cal B}^{e  g\to e Q \bar Q}\,  ,
\label{eq:BQQb}
\end{equation}
with
\begin{equation}
 {\cal B}^{e  g\to e Q\bar Q} = \frac{1}{2}\, \frac{z (1-z)}{D^3}\, \left
  (1-\frac{M_Q^2}{M_\perp^2} \right ) a(y)  
\left \{ \big [2\, z (1-z) \, b(y) - 1\big ]
  \frac{Q^2}{M_\perp^2} +2 \,\frac{M_Q^2}{M_\perp^2}\right \}~,
\label{eq:BQQb2}
\end{equation}
$D \equiv D \left (z,Q^2/M_\perp^2 \right ) = 1 + z (1-z) Q^2/M_\perp^2$,
$a(y) = 2 -y (2-y)$,  
$b(y) =  [6 -y (6-y)]/a(y)$.

\begin{figure}[t]
 \includegraphics[angle=0,width=0.5\textwidth]{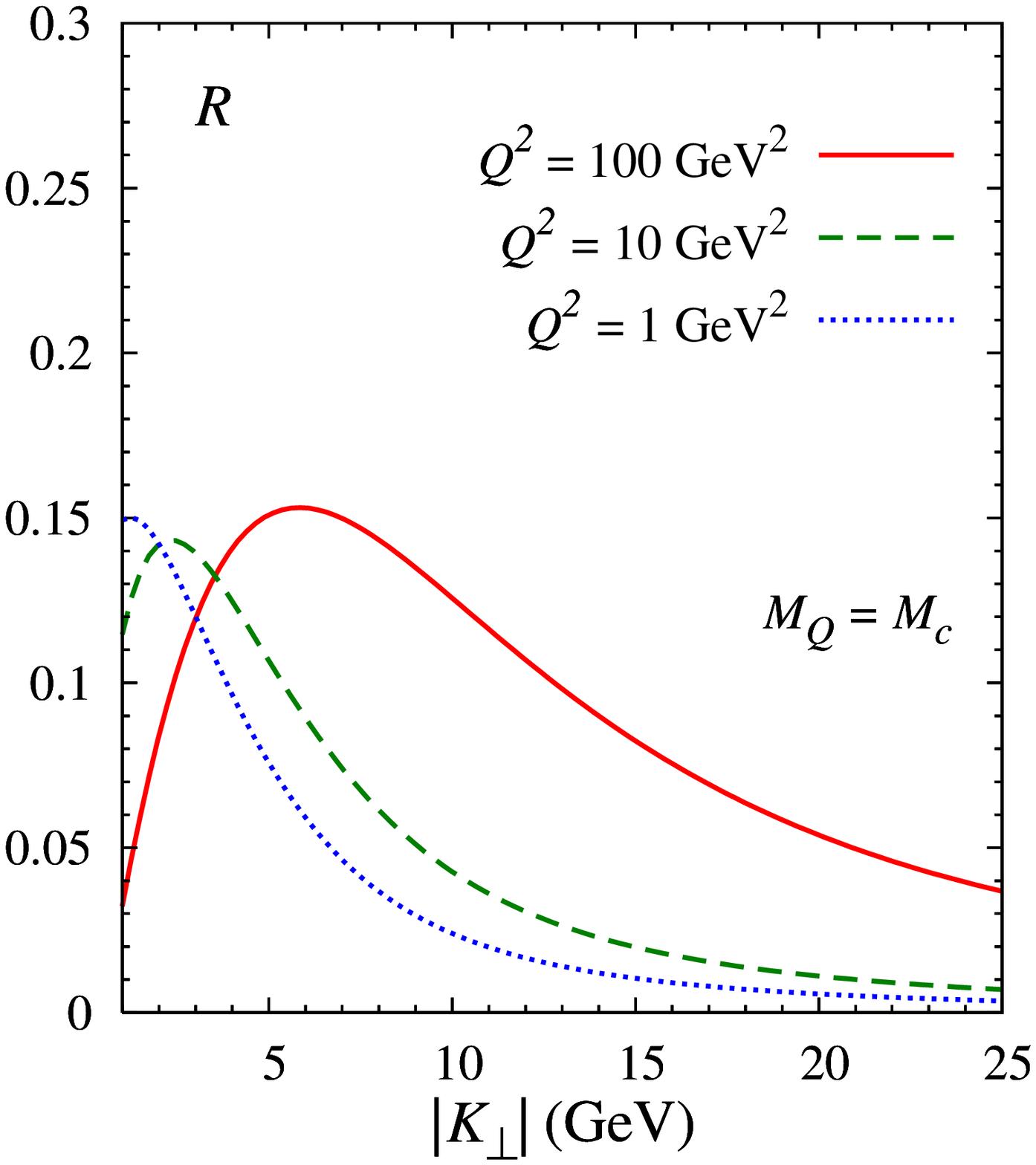}
 \includegraphics[angle=0,width=0.5\textwidth]{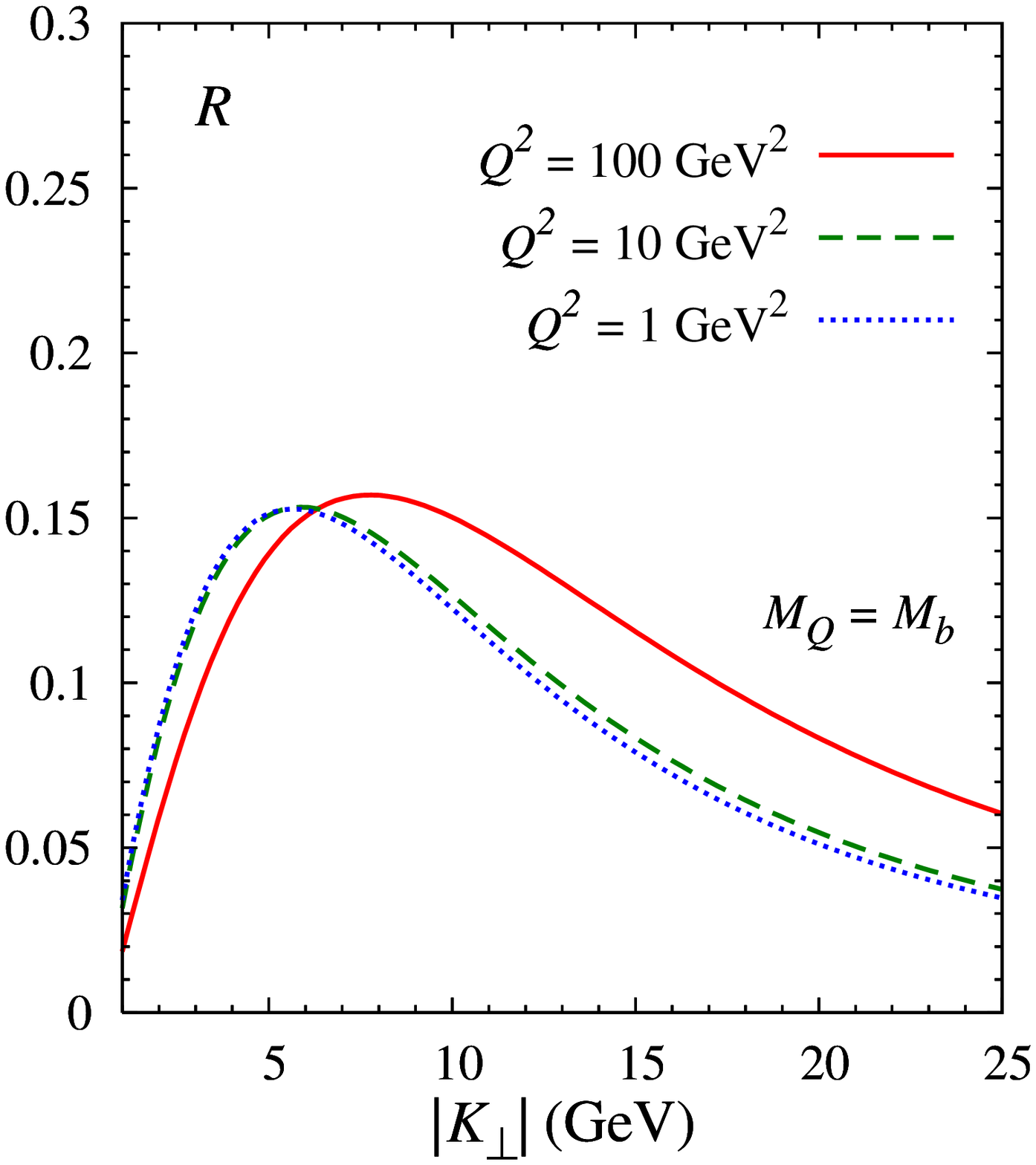}
 \caption{Upper bound of $\vert\langle \cos 2 (\phi_{\sT}-\phi_\perp)\rangle\vert$ defined in Eq.\ \eqref{BBMPeq:bound} as a function of $\vert \bm K_\perp \vert$ at different values of $Q^2$, with $y=0.01$ and $z=0.5$. 
\label{BBMPfig:asy} }
\end{figure}

Since $h_1^{\perp\, g}$ is completely unknown, we estimate the maximum
asymmetry that is allowed by the bound 
\begin{equation}
|h_1^{\perp\, g (1)}(x)| \leq  f_1^g(x)\, ,
\label{BBMPbound}
\end{equation}
where the superscript $(1)$ denotes the $n=1$ transverse moment (defined as
$f^{(n)}(x) \equiv
\int  d^2 \bm{p}_{\sT}\;
\left(\bm{p}_{\sT}^2/2 M^2\right)^n
\;f(x, \bm{p}_{\sT}^2)$).
The function $R$, defined as the upper bound of the absolute value of 
$\langle \cos 2 (\phi_{\sT}-\phi_\perp) \rangle$,
\begin{equation}
\vert\langle \cos 2 (\phi_{\sT}-\phi_\perp)\rangle\vert \equiv
\left| \frac{\int d^2 \bm{q}_{\sT}
\, \cos 2 (\phi_{\sT}-\phi_\perp) \, d\sigma}{\int d^2 \bm{q}_{\sT}
\, d\sigma}\right| =  \frac{\int d\bm{q}_{\sT}^2\,  \bm{q}_{\sT}^2\, | B |}{
2 \int d \bm{q}_{\sT}^2 \, A}
\le   \frac{| {\cal B}^{e  g\to e Q\bar Q}  |}{{\cal{A}}^{e  g\to e Q\bar Q}} \equiv R\, ,
\label{BBMPeq:bound}
\end{equation}
is depicted in Fig.\ \ref{BBMPfig:asy} as a function of $\vert \bm K_\perp\vert$ ($>$ 1 GeV) at different values of 
$Q^2$ for charm (left panel) and bottom (right panel)
production. We have selected $y= 0.01$, $z=0.5$, and taken $M_c^2=$ 2 
 GeV$^2$, $M_b^2=$ 25 GeV$^2$. 
Such large asymmetries
would probably allow an extraction of $h_1^{\perp\, g}$ at EIC (or LHeC).

\begin{figure}[t]
 \includegraphics[angle=0,width=0.5\textwidth]{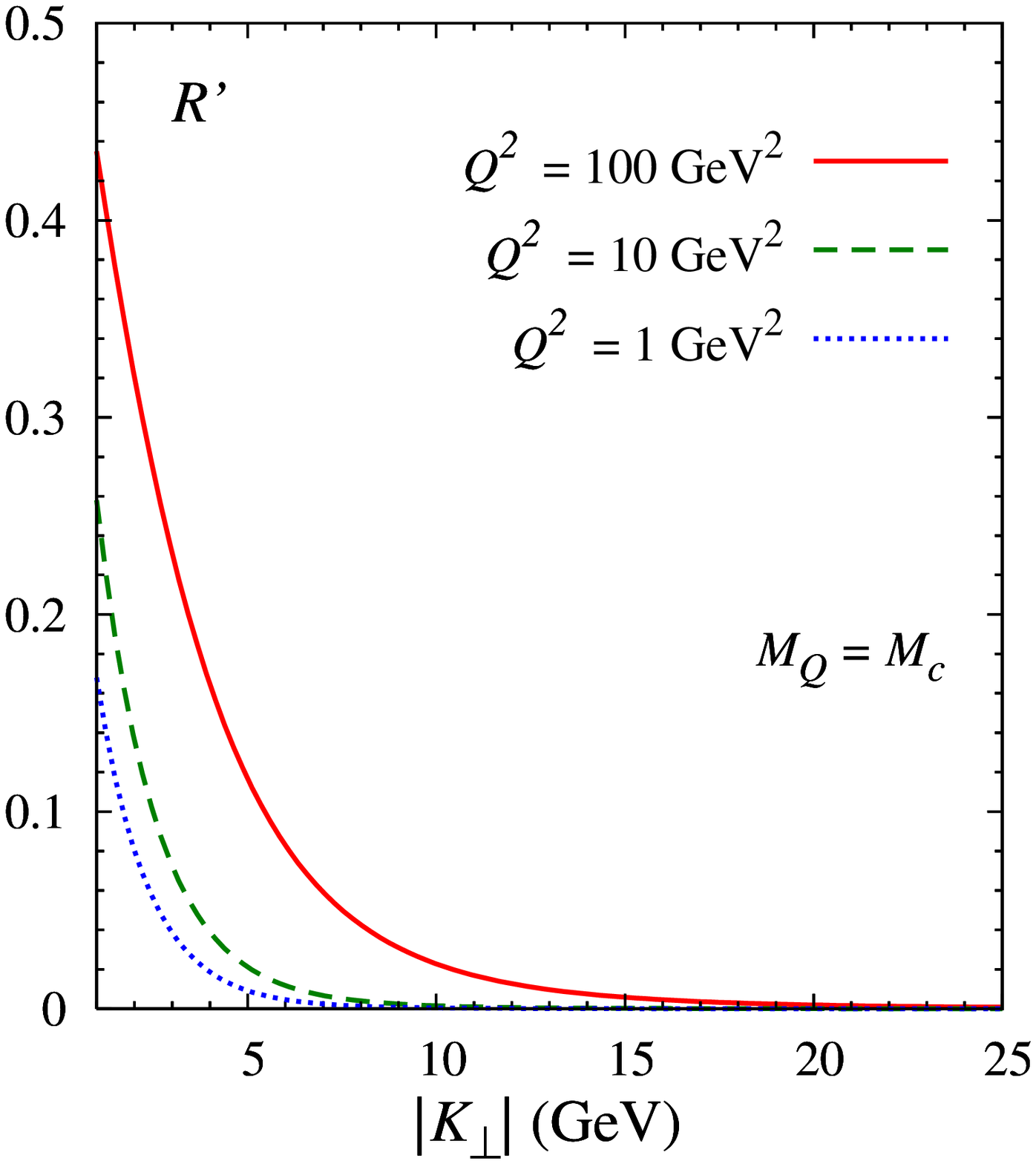}
 \includegraphics[angle=0,width=0.5\textwidth]{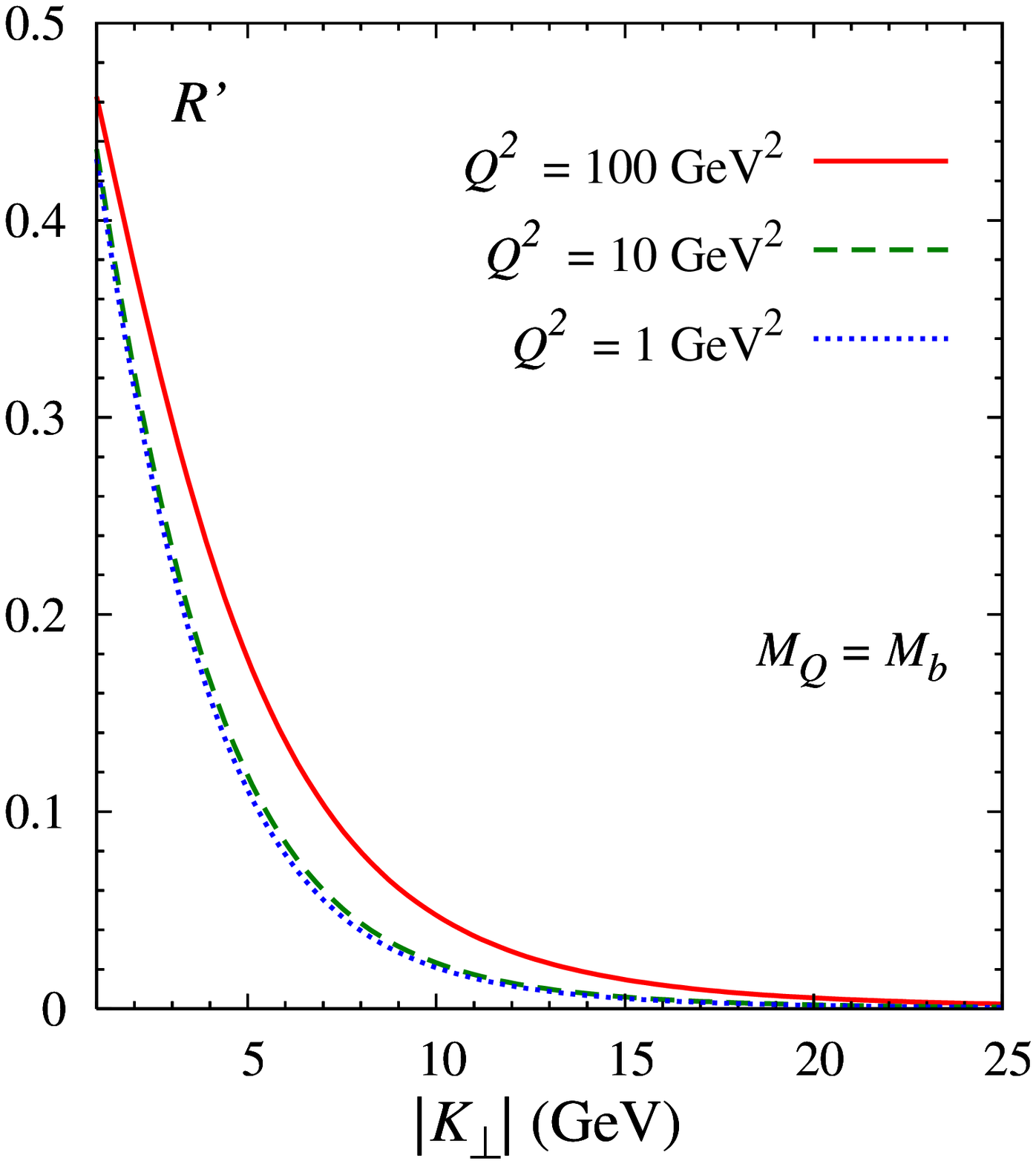}
 \caption{Same as in Fig.\ \ref{BBMPfig:asy}, but for the upper bound $R'$ 
of  $\vert \langle \cos 2 (\phi_\ell-\phi_{\sT}) \rangle \vert$.
\label{BBMPfig:asyp}} 
\end{figure}

If one keeps the lepton plane angle $\phi_\ell$, there
are other azimuthal dependences, such as a $\cos 2(\phi_\ell - \phi_{\sT})$. The bound on $\vert \langle \cos 2(\phi_\ell - \phi_{\sT}) \rangle   \vert $, 
denoted as $R'$,
is shown in Fig.\ \ref{BBMPfig:asyp} in the same kinematic
region as in Fig.\ \ref{BBMPfig:asy}. One can see 
that $R'$ can be larger than $R$, but only at smaller $\vert \bm K_\perp\vert$. $R'$ falls off more rapidly at larger 
values of $\vert \bm  K_\perp\vert$ than $R$. We note that it is essential that 
the individual transverse momenta $K_{i\perp}$ are reconstructed with an accuracy $\delta K_\perp$
better than the magnitude of the sum of the transverse momenta 
$K_{1\perp}+K_{2\perp}=q_{\sT}$. This means one has to satisfy 
$\delta K_\perp \ll \vert q_{\sT}\vert \ll \vert K_\perp\vert$,
which will require a minimum $\vert K_\perp\vert$.

The cross section for the process 
$e\, h\to e^\prime\,{\rm jet}\, {\rm jet}\, X$
can be calculated in a similar way and is analogous to Eq.\ 
\eqref{BBMPeq:cso}. In particular, the explicit expression for $B$ can be obtained
from the one for HQ production taking $M_Q=0$, 
while $A$ now depends also on $x_{\scriptscriptstyle B}$ 
and receives a contribution from the subprocess $\gamma^* q \to g q$ as well,
not just from  $\gamma^* g \to q \bar{q}$. Therefore, the maximal asymmetries (not shown) are  
smaller than for HQ pair production. 


\section{Conclusions}

Studies of the azimuthal asymmetry of jet or heavy quark pair production 
in $e \,p$ 
 collisions 
can directly probe $h_1^{\perp\, g}$, the distribution of linearly polarized 
gluons inside 
unpolarized hadrons.
Breaking of TMD factorization is expected in $p\,p$ or $p \,\bar{p}$ 
collisions, hence 
a comparison between extractions from these two types of processes 
would  clearly signal the dependence on ISI/FSI. 
 The contribution of  $h_1^{\perp\, g}$
to diphoton
production has also  been studied \cite{Qiu:2011ai}.
Since the proposed measurements are relatively simple 
(polarized beams are not required), we believe  that the experimental 
determination of $h_1^{\perp\, g}$ 
 and the analysis of its potential process dependence will be feasible 
in the future. 


\begin{theacknowledgments}
C.P.~is supported by Regione Autonoma della Sardegna (RAS) through a research grant under the PO Sardegna FSE 2007-2013, L.R. 7/2007. 
This research is part of the FP7 EU-programme Hadron Physics (No.\ 227431). SLAC-PUB-14494.
\end{theacknowledgments}



\bibliographystyle{aipproc}   


\begin{thebibliography}{9}
\bibitem{Mulders:2000sh}
  P.J.~Mulders and J.~Rodrigues,
  Phys.\ Rev.\  D {\bf 63}, 094021 (2001).

\bibitem{Boer:2009nc}
  D.~Boer, P.J.~Mulders and C.~Pisano,
  Phys.\ Rev.\ D {\bf 80},  094017 (2009).

\bibitem{Boer:2010zf}
  D.~Boer, S.J.~Brodsky, P.J.~Mulders and C.~Pisano, 
  Phys.\ Rev.\  Lett.\ {\bf 106}, 132001  (2011).

\bibitem{Boer:1997nt}
  D.~Boer and P.J.~Mulders,
  Phys.\ Rev.\  D {\bf 57}, 5780  (1998).

\bibitem{Rogers:2010dm}
  T.C.~Rogers and P.J.~Mulders,
  Phys.\ Rev.\  D {\bf 81}, 094006 (2010). 

\bibitem{Boer:2007nd}
  D.~Boer, P.J.~Mulders and C.~Pisano,
  Phys.\ Lett.\  B {\bf 660}, 360 (2008).





\bibitem{Qiu:2011ai}
  J.~Qiu, M.~Schlegel and W.~Vogelsang,
  arXiv:1103.3861 [hep-ph].


\end{thebibliography}



\end{document}